\documentclass[12pt]{iopart}

\expandafter\let\csname equation*\endcsname\relax
\expandafter\let\csname endequation*\endcsname\relax
\usepackage{amsmath, amssymb}
\usepackage{graphicx}
\usepackage{comment}
\usepackage{hyperref}
\usepackage{soul,color}
\usepackage{braket, xfrac}

\newcommand{\up}{\uparrow}
\newcommand{\down}{\downarrow}
\newcommand{\average}[1]{\left\langle{#1}\right\rangle}
\newcommand{\adag}{a^{\dagger}}
\newcommand{\gdn}{\Gamma_{\down}}
\newcommand{\gup}{\Gamma_{\up}}

\newcommand{\sz}{\average{\sigma^z}}
\renewcommand{\sp}{\average{\sigma^+}}
\newcommand{\sm}{\average{\sigma^-}}
\renewcommand{\aa}{\average{a a}}
\newcommand{\ada}{\average{\adag a}}

\newcommand{\zz}{C^{zz}}

\newcommand{\pp}{C^{++}}
\renewcommand{\pm}{C^{+-}}

\newcommand{\ap}{C^{a+}}
\newcommand{\am}{C^{a-}}

\newcommand{\new}[1]{#1}

\let\Re\relax \DeclareMathOperator\Re{\mathrm{Re}}%
\let\Im\relax \DeclareMathOperator\Im{\mathrm{Im}}%

\begin{document}

\title{Superradiant and Lasing States in Driven-Dissipative Dicke Models}
\author{Peter Kirton and Jonathan Keeling}
\address{SUPA, School of Physics and Astronomy, University of St Andrews, St Andrews, KY16 9SS, United Kingdom}

\date{\today}

\begin{abstract}
	We present the non-equilibrium phase diagram of a model which can demonstrate both  Dicke--Hepp--Lieb superradiance and regular lasing by varying the coherent and incoherent driving terms. We find that the regions in the phase diagram corresponding to superradiance and standard lasing are always separated by a normal region. We analyse the behaviour of the system using a combination of exact numerics based on permutation symmetry of the density matrix for small to intermediate numbers of molecules, and second order cumulant equations for large numbers of molecules. We find that the nature of the photon distribution in the superradiant and lasing states are very similar, but the emission spectrum is very different.  We also show that in the presence of both coherent and incoherent driving, a period-doubling route to a chaotic state occurs.
\end{abstract}
\maketitle

\section{Introduction}

When large numbers of emitters couple to light, there can be constructive interference effects which enhance both emission and coupling to light.  Such enhancement can arise either from the bosonic stimulation due to occupation of the photon mode, or from constructive interference between emission pathways in the atoms, or both.  These effects have consequences both for the dynamics of an initially excited state, and for the possibility to maintain steady states with coherent matter and light. The term \textit{superradiance} is used in the literature to describe a variety of these phenomena, which are related, but distinct. The original use of the term superradiance by Dicke~\cite{Dicke1954} described the enhanced transient emission from a cloud  of initially excited atoms when the size of the cloud is smaller than the wavelength of the emitted light. Later, it was noted that a related model, describing many atoms placed in a single mode cavity, can undergo a ground state phase transition from a normal state at weak light-matter coupling to a superradiant (SR) state with a macroscopically occupied cavity mode~\cite{Hepp1973,Wang1973,Carmichael1973}. However, the existence of this phase transition in the ground state of real systems is under debate due to the effects of \new{diamagnetic} terms in the minimal coupling Hamiltonian~\cite{Rzazewski1975,Nataf2010,Viehmann2011,Vukics2012,Bamba2014,Vukics2014,Jaako2016}.   

In spite of the debate regarding the ground state phase transition, a related phase transition certainly is achievable using a Raman driven scheme~\cite{Dimer2007, Ritsch2013}, which has been realised in recent experiments with ultracold atoms in optical cavities~\cite{Baumann2010, Baumann2011, Baden2014, Klinder2015}. In such a scheme, two low-lying states of the atom can be considered to form an effective two-level system.  These levels are then connected by Raman transitions, via virtually populated excited levels of the atoms.  The Raman scheme involves both light in the cavity and an external pump.  This means that the
effective coupling between atoms and the cavity mode can be controlled by the strength of external pump, circumventing the issues with the ground state transition.  This driven dissipative system is able to show a transition very similar to the ground state transition described above. 
 These experiments prompted much theoretical investigation~\cite{Nagy2010, Keeling2010,Oeztop2012, Bhaseen2012, Torre2013, Piazza2013, Nagy2015, Torre2016, Kirton2017, Gelhausen2017} into the nature of the phase transition.

There are many similarities between the steady state superradiance transition
in the Raman driven context described above and the transition seen in simple models of a two-level laser~\cite{Scully1997}. Both involve a transition from a state with an empty cavity to a macroscopically occupied photon mode, and
both operate due to external driving balancing cavity loss. Microscopically, the main difference between them is in the mechanism by which the driving occurs. In a two-level description of a laser the upper level is populated via an \textit{incoherent} pumping process, while in the Raman driven Dicke model the driving appears via ``counter-rotating'' terms in the Hamiltonian, which provide a \textit{coherent} pump.  It is worth noting in this context that the superradiant laser discussed in~\cite{Bohnet2012, Bohnet2012a, tieri2017} corresponds to an incoherent pumping process~\cite{tieri2017}, and is distinguished from a standard laser by operating deep in the bad cavity limit of cavity QED.

In this paper we look at the ways in which these coherent and incoherent driving processes can interact and lead to a rich phase diagram with regions showing differing behaviours. By studying a model which is able to show both steady-state superradiance and standard two-level lasing, we may study whether these limiting behaviours can be continuously connected, and whether other forms of lasing or superradiance exist when both driving processes are combined.  We find that the phase diagram as a function of both coherent and incoherent pumping shows two distinct phases where the photon mode is macroscopically occupied. These are continuously connected to the SR phase and the lasing phase. The lasing phase can only occur in the region where the spins are inverted and the SR phase below inversion. We go on to explore the behaviour in these two regions using using a second order cumulant expansion which is valid at large (but finite) $N$, backed up by exact numerics based on permutation symmetry at intermediate $N$. We find 3 distinct parameter regions: when the coherent pumping term is small we find that the system transitions to a lasing state at large pumping. When the coherent pumping is large there is a superradiant state which exists only when the incoherent pump is weak. Between these two limits a crossover is seen where the system goes superradiant-normal-lasing as the incoherent pump power is increased.

The structure of this paper is as follows. In Sec.~\ref{sec:model} we present the model we use and give some motivation for how it could be experimentally realised. In Sec.~\ref{sec:MF} we give the mean field equations and show how these can be used to calculate the stability of the normal state. Then in Sec.~\ref{sec:cumulant} we go beyond these mean field \new{equations} to derive the second order cumulant equations which allow us to examine second order moments of the distribution. We then go on, in Sec.~\ref{sec:exact} to show how these results compare to exact solutions using permutation symmetric methods, these also allow us to calculate higher moments such as the Fano factor and probability distribution for occupation of the photon mode through both the SR and lasing transitions. In Sec.~\ref{sec:spectrum} we calculate the emission spectrum of the model, finding a distinct qualitative difference between the behaviours in the lasing and SR regions of the phase diagram. In Sec.~\ref{sec:inverted} we show that a deeper understanding of the relation between lasing and superradiant states can be reached by  considering the phase diagram as a function of the frequency of the photon mode.  This reveals that the model actually has four distinct phases: the normal SR and lasing phase which typically occur for positive photon frequency but also an inverted version of each state which occur when the photon is inverted. This also leads us to uncover some interesting chaotic dynamics which occur in the mean field and cumulant equations. Finally, in Sec.~\ref{sec:conclusions} we present our conclusions. 

\section{Model}
\label{sec:model}

\begin{figure}
\centering
 \includegraphics[width=0.75\columnwidth]{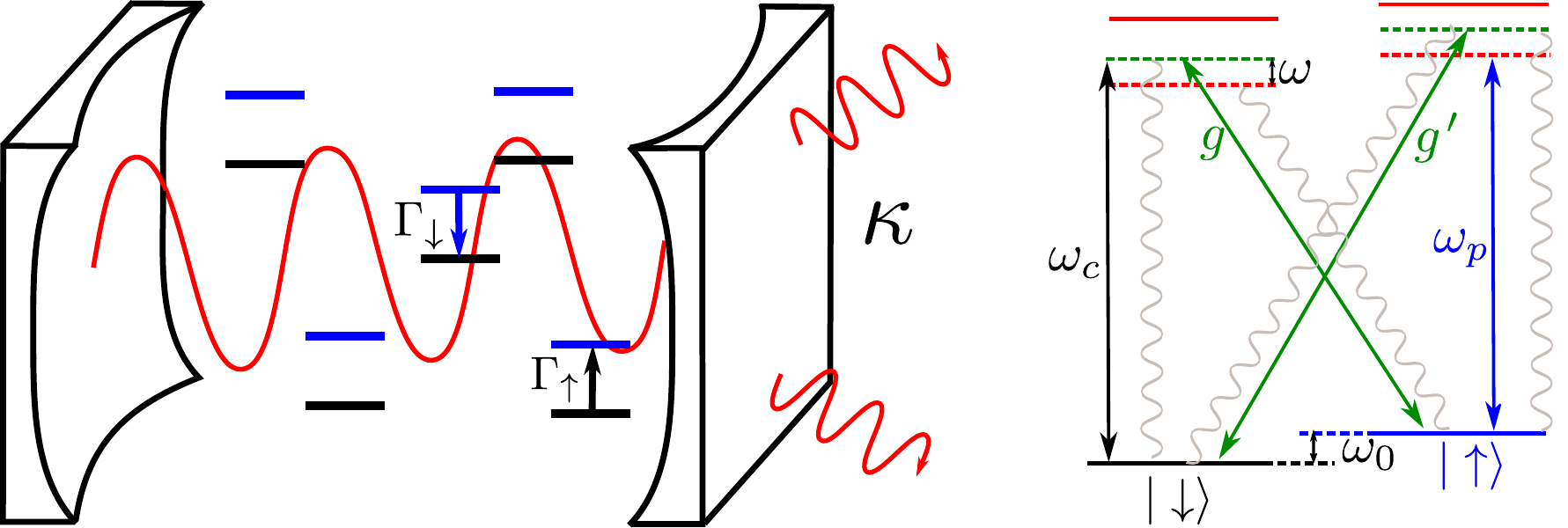}
 \caption{\label{fig:schematic} Schematic diagram of the model we consider. Left: cartoon showing the loss processes involved. Right: example energy level diagram for the four level system. The two levels which remain after adiabatic elimination are the lower black and blue levels.}
\end{figure}

We examine a model which can show both Dicke superradiance and a standard lasing transition. The model has the Hamiltonian
\begin{equation} \label{eqn:DickeHam}
	H = \omega a^\dagger a + \sum_i^N\frac{\omega_0}{2} \sigma^z_i + g(\sigma_i^+a+\sigma^-_i\adag) 
	+ g'(\sigma_i^-a+\sigma^+_i\adag),
\end{equation}
which describes a single photon mode (annihilation operator $a$) interacting with an ensemble of two level atoms (described by the Pauli operators $\sigma_i$). A schematic diagram showing a possible realisation of this model in a cold atom setting is shown in Fig.~\ref{fig:schematic} along with the four level scheme which can produce this model after adiabatic elimination of the upper (detuned) levels~\cite{Dimer2007}. 
  
This scheme allows us to separate the co- and counter-rotating terms of the light-matter interaction~\cite{Dimer2007} to have prefactors $g$ and $g'$ respectively. The master equation for the density matrix, $\rho$, of the system is given by:
\begin{equation} \label{eqn:ME}
	\frac{d \rho}{dt} = -i[H,\rho] + \kappa \mathcal{D}[a] + \sum_i^N \gdn \mathcal{D}[\sigma^-_i]+\gup \mathcal{D}[\sigma^+_i],
\end{equation}
which includes photon loss at rate $\kappa$, incoherent loss of atomic excitations at rate $\gdn$ and an incoherent pumping process at rate $\gup$ as standard Lindblad terms with $\mathcal{D}[x]=x\rho x^\dagger- 1/2\{x^\dagger x, \rho \}$.
We note that both the atomic pumping and dissipation terms could in principle be engineered using the same Raman scheme as illustrated in Fig.~\ref{fig:schematic}, but not involving the cavity --- i.e. spontaneous Raman processes involving real emission from excited states of the atoms.  
In most experiments~\cite{Baumann2010, Baumann2011, Baden2014, Klinder2015}, such terms are deliberately suppressed, by using the fact that the strength of this incoherent Raman process scales as the inverse square of the laser detuning from the atomic resonance, while the coherent terms scale only as the inverse of the detuning.  In  general, allowing real decay also induces additional dephasing terms $\mathcal{D}[\sigma^z_i]$ which have been discussed elsewhere~\cite{Torre2016,Kirton2017}.  Here, for simplicity, we neglect additional dephasing, and focus on the competition between incoherent and coherent driving processes.

It is clear that when $g=g'$, $\gup=0$ this is the regular open Dicke model which we have previously studied~\cite{Kirton2017}. This model has a phase transition to a SR state which spontaneously breaks a $\mathbb{Z}_2$ symmetry which exists since the model is invariant under the exchange $a\to -a$, $\sigma_i^x \to -\sigma_i^x$. The presence of the two loss terms $\kappa$ and $\gdn$ modify the phase transition by slightly shifting the critical point to~\cite{Torre2016, Kirton2017, Gelhausen2017}
\begin{equation}
\label{eqn:gcfull}
g_c^2N = \frac{1}{4\omega_0\omega}\left(\omega_0^2+\frac{\gdn^2}{4} \right)\left(\omega^2+\frac{\kappa^2}{4}\right).
\end{equation}

In contrast, if $g'=0$ then this shows a regular lasing transition at a particular value of $\gup$ which for small $\kappa$ is when $\gup\simeq \gdn$. This transition breaks a $U(1)$ symmetry in which the \new{equations} are invariant under the transformation $a \to a \text{e}^{i\phi}$, $\sigma^-_i \to \sigma^-_i \text{e}^{i\phi}$. 

By varying $g'$ and $\gup$, the behaviour can be continuously varied between the two cases described above. This allows us to understand how the physics crosses over between these two limits, and to identify probes to experimentally distinguish between the lasing and superradiance that both occur in this model. Our aim in the remainder of this paper will be to study this model with increasing levels of sophistication, in order to identify the phases that occur, and to characterise the behaviour in each phase.

\section{Mean-field stability analysis}
\label{sec:MF}

We start our analysis by looking at the mean-field equations for the cavity and the spins.  As demonstrated in previous work~\cite{Kirton2017}, even with dephasing and dissipation, the mean-field analysis gives a reasonable picture of the states that can arise.  This decomposition  requires us to break correlations between the different parts of the system at first order, resulting in the following set of equations,
\begin{gather}
  \label{eq:mfa}
  \partial_t \average{a} = -\left(i\omega+\frac{\kappa}{2}\right)\average{a} - iN\Bigl(g\sm +g'\sm^\ast\Bigr), \\
  \label{eq:mfb}
  \partial_t \sm = -\left(i\omega_0+\frac{\Gamma_T}{2}\right)  \sm +i\Bigl(g\average{a} +g'\average{a}^*\Bigr)\sz, \\  
  \label{eq:mfc}
  \partial_t \sz = 4g\Im\left[\average{a}\sm^\ast\right] - 4g'\Im\left[\average{a}\sm\right]  -\Gamma_T\sz +\gup-\gdn,
\end{gather}
where $\Gamma_T=(\gdn+\gup)$. When $g'=0$ these are just the Maxwell-Bloch equations for a laser~\cite{Haken1970}. 

\begin{figure}
\centering
 \includegraphics[width=0.8\columnwidth]{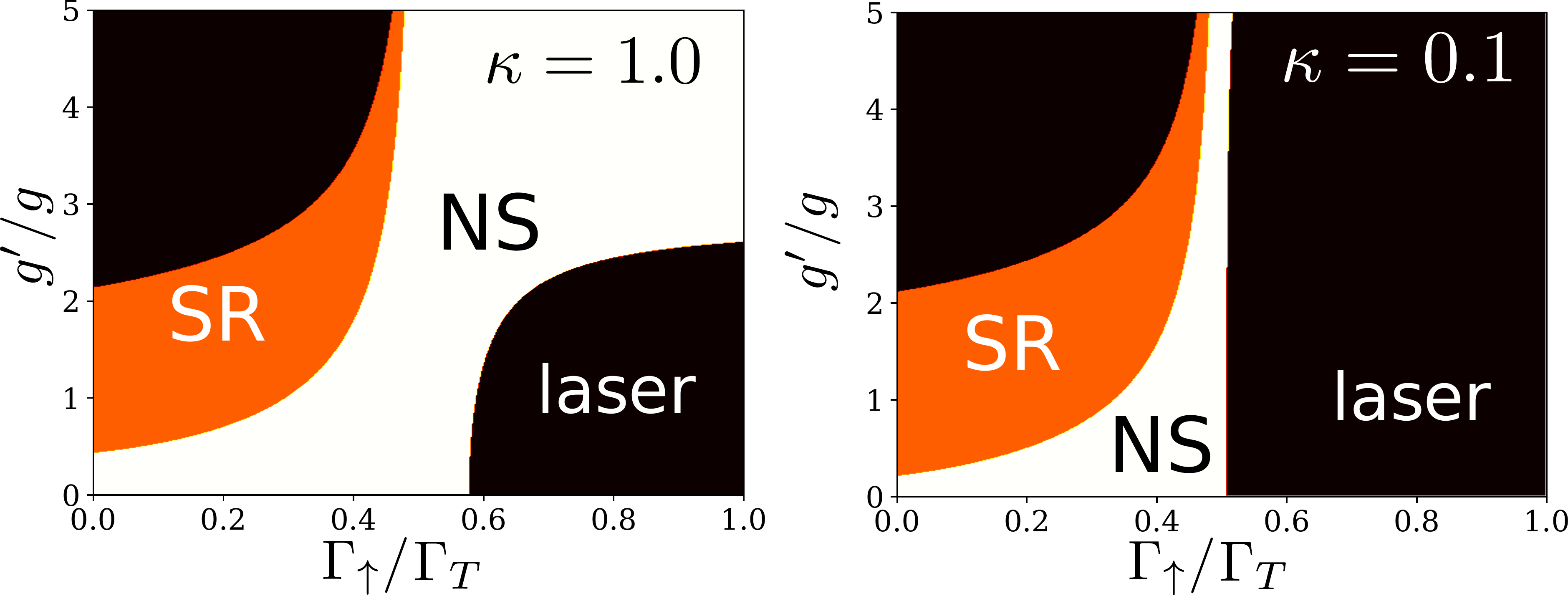}
 \caption{\label{fig:MF} Mean field stability of the normal state for different values of photon loss, $\kappa$. In the white region the normal state is stable, while in the orange and black regions the stability matrix has 1 and 2 unstable eigenvalues respectively.  Parameters are  $\omega=1$, $g=0.9$, $\Gamma_T=0.5$ and $\kappa$ as indicated. All  quantities are in units of $\omega_0$. \new{The normal state  (NS) has a vanishing photon population, while the superradiant (SR) and lasing states have a macroscopically occupied photon mode.}}
\end{figure}

The normal state  has $\average{a}_{ns}=\sm_{ns}=0$ and $\sz_{ns}=(\gup-\gdn)/\Gamma_T$. Linearising around this solution with $\delta a = \average{a} - \average{a}_{ns}$, $\delta s = \sm - \sm_{ns}$  etc.\ gives the following matrix
\begin{equation}
\frac{d}{dt}\begin{pmatrix}\delta a \\ \delta a^* \\ \delta s \\ \delta s^*\end{pmatrix} = 
\begin{pmatrix} 
  -i\omega -\kappa/2 & 0 & -ig & -ig' \\
  0 & i\omega -\kappa/2 & ig' & ig \\
  ig\sz_{ns} & ig'\sz_{ns} & -i\omega_0-\Gamma_T/2 & 0 \\
  -ig'\sz_{ns} & -ig\sz_{ns} & 0 &  i\omega_0-\Gamma_T/2
\end{pmatrix}
\begin{pmatrix}
  \delta a \\ \delta a^* \\ \delta s \\ \delta s^*
\end{pmatrix}.
\end{equation}
This $4\times 4 $ matrix equation arises since $\average{a}$ couples both $\sp$ and its conjugate $\sm$. We may ignore the equation of motion for $\sz$ since this only couples to itself in the linearised system, and so does not contribute to the stability analysis.

When the real part of (at least) one of the eigenvalues of this matrix becomes positive the normal state is unstable to perturbations and the photon mode acquires a macroscopic occupation. This linear stability approach allows us to very quickly explore large regions of parameter space. In Fig.~\ref{fig:MF} we explore this phase diagram as a function of both the coherent pumping term $g'$ and the incoherent pumping term $\gup$. In order to show the full range of $\gup/\gdn$ (i.e. from $0$ to $\infty$) we vary the incoherent pump while keeping the total decay of the spins, $\Gamma_T$, constant, so that $\gup/\Gamma_T=1$ corresponds to $\gup/\gdn=\infty$.

We see that there are two distinct regions in which the normal state is unstable, one, at small values of $\gup$, is continuously connected to the SR region  while the other, at small $g'$, is connected to the lasing transition. We label these regions ``SR'' and ``laser'' respectively on the phase diagram. We see that, even when the photon loss rate is very small, these two regions are always separated by a (possibly small) normal region which is always present exactly at the point where the spins become inverted $\gup=\gdn$. The main change as $\kappa$ is reduced is to the shape of the lasing region which we see move towards the inversion line as the losses are reduced. We note that the lasing region can never cross this line from above while the SR region never crosses it from below.

In Fig.~\ref{fig:MF}, for the regions where the normal state is unstable, we also indicate whether there exists only a single unstable mode, or a pair of unstable modes.  In the latter case, this corresponds to a complex conjugate pair of eigenvalues, and describes a mode which oscillates as well as becoming unstable.  As we will discuss further below, this is exactly as expected for the laser, as the frequency of the lasing mode generally matches the cavity frequency, and so the lasing solution is expected to be time dependent (oscillatory) in the frame we are working in.  This means that the instability to lasing should involve  an eigenvalue with an imaginary part, corresponding to the lasing frequency. Since eigenvalues come in complex conjugate pairs, the instability in this case is a Hopf bifurcation, where a complex conjugate pair of eigenvalues simultaneously become unstable.  In contrast, the ``standard'' superradiant state is expected to be stationary, and the instability to it thus corresponds to a mode which grows without oscillating, thus a single mode is unstable at the phase boundary.  The phase diagram observed is compatible with this observation, however more complicated behaviour will be seen below.

\section{Second order cumulant expansion}
\label{sec:cumulant}

While mean-field theory is useful to determine the structure of the phase
diagram, it can only capture the behaviour in the limit $N \to \infty$. As
such, it is not always possible to directly compare mean-field
theory results to exact results at finite $N$. We can go beyond mean-field theory by finding the equations for all the second moments of the photon and atomic distributions. This approach allows us to describe quantities such as the photon number accurately and without the need to explicitly break symmetries in the initial conditions.  Indeed, in writing these equations we will choose to keep only those terms that respect the symmetries in the problem.  This approach is directly analogous to the semiclassical approach used in laser theory~\cite{Haken1970}, which considers equations of motion for the photon number, while incorporating both spontaneous and stimulated processes.

The second moments of the photons obey the equations of motion
\begin{align}
  \partial_t\ada &= -\kappa\ada - 2N(g\Im[\ap]+g'\Im[\am]),  \\
  \partial_t\aa &= -(2i\omega+\kappa)  \aa - 2iN(g\am+g'\ap),
\end{align}
while the photon matter correlations are
\begin{gather}
  \begin{split}
  \partial_t\ap = 
  -\left(i\omega+\frac{\kappa}{2}+\tilde\Gamma\right)\ap + i\omega_0 \ap 
  \\ -ig\left[\left(N-1\right)\pm + \sz\ada +\sfrac{1}{2}(1+\sz) \right] \\
      -ig'\left[\left(N-1\right)\pp + \sz\aa\right],
  \end{split}
  \\
  \begin{split}
  \partial_t\am = 
  -\left(i\omega+\frac{\kappa}{2}+\tilde\Gamma\right)\am - i\omega_0 \am 
  \\ 
  -ig\left[\left(N-1\right)(\pp)^* - \sz\aa\right] \\
  -ig'\left[\left(N-1\right)\pm - \sz\ada +\sfrac{1}{2}(1-\sz) \right],
  \end{split}
\end{gather}
here e.g.\ $\ap=\average{a\sigma^+_i}$. 
There are also matter-matter correlations given by
\begin{gather}
  \partial_t\pp = 2i\omega_0\pp - 2\tilde\Gamma\pp -2i\sz(g(\am)^*+g'\ap), \\
  \begin{split}
  \partial_t\zz = 8\sz(g\Im[\ap] - g'\Im[\am])\\ - 2\gdn(\zz +\sz) - 2\gup(\zz -\sz),
  \end{split}
  \\
  \partial_t\pm = -2\tilde\Gamma\pm - 2\sz(g\Im[\ap]-g'\Im[\am]).  
\end{gather}
where e.g.\ $\pm=\average{\sigma^+_i\sigma^-_{j\neq i}}$ gives the correlation between $\sigma^+$ at site $i$ and $\sigma^-$ at a different site $j$.  We also need the equation for $\sz$, which does not break the symmetry, and takes a form similar to that in Eqn.~\eqref{eq:mfc} but without breaking the second order correlations,
\begin{equation}
	  \partial_t \sz = 4g\Im\left[C^{a+}\right] - 4g'\Im\left[C^{a-}\right]  -\Gamma_T\sz +\gup-\gdn.
\end{equation}
These reproduce the equations in Ref.~\cite{Kirton2017} when $g'=g$ and $\gup=0$. 

\begin{figure}
\centering
 \includegraphics[width=1.0\columnwidth]{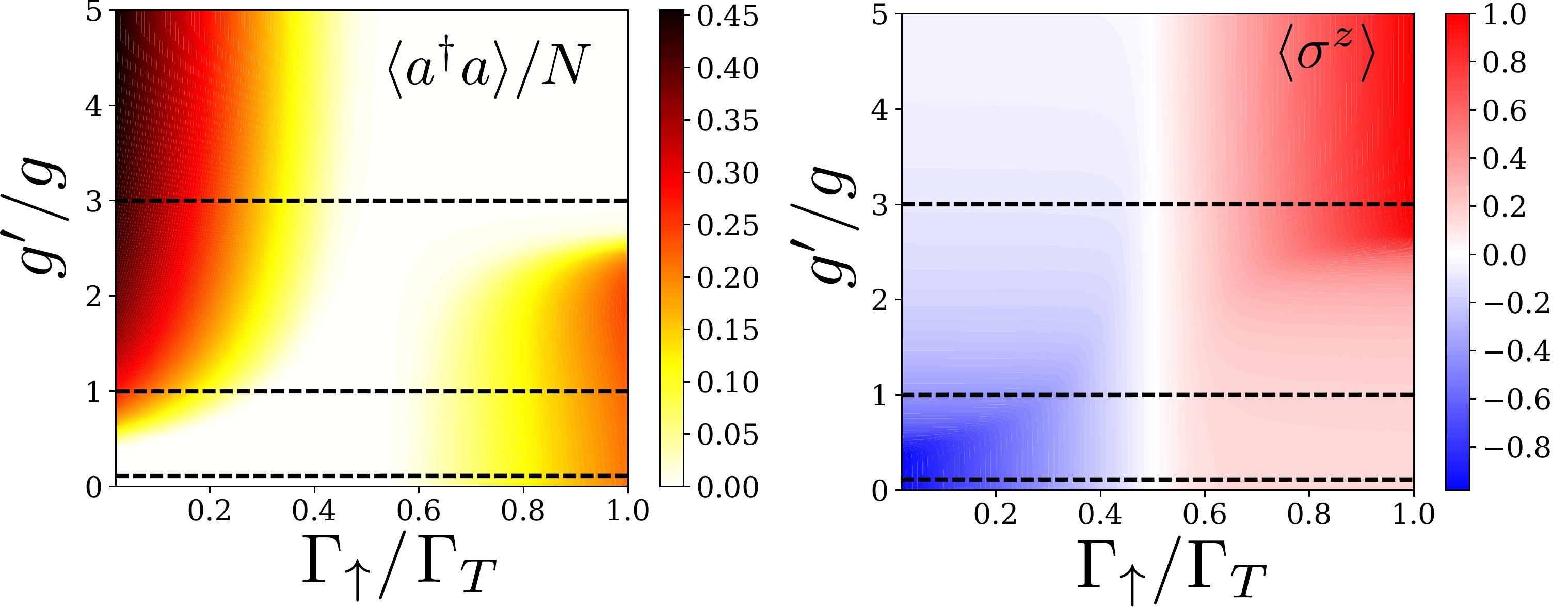}
 \caption{.\label{fig:Gup_gp} Phase diagram of (a) photon number and (b) spin inversion vs $\gup$ and $g'$ calculated using the second order cumulant equations. The parameters are the same as in Fig.~\ref{fig:MF} (a) with $N=500$. 
   }
\end{figure}

In Fig.~\ref{fig:Gup_gp} we show how these equations can be used to examine the phase diagram which was plotted in Fig.~\ref{fig:MF}. We may check that the region with a macroscopic photon number is in very good agreement with that predicted by mean field theory, with the SR dome at large $g'$ and small $\gup$ and the lasing region at large $\gup$ and small $g'$. By looking at the spin state we see directly that the SR region is where the spins are not inverted and the lasing region corresponds to spin inversion. The vertical white region in Fig.~\ref{fig:Gup_gp}(b) shows exactly where this spin inversion occurs. This is always at the point where $\gup=\gdn$ (at least up to corrections of order $\kappa/N$) since here the photons have no occupation and so the dynamics are purely determined by the driven atoms.

\section{Comparison to exact results and higher order correlations}
\label{sec:exact}

\begin{figure}
\centering
 \includegraphics[width=1.0\columnwidth]{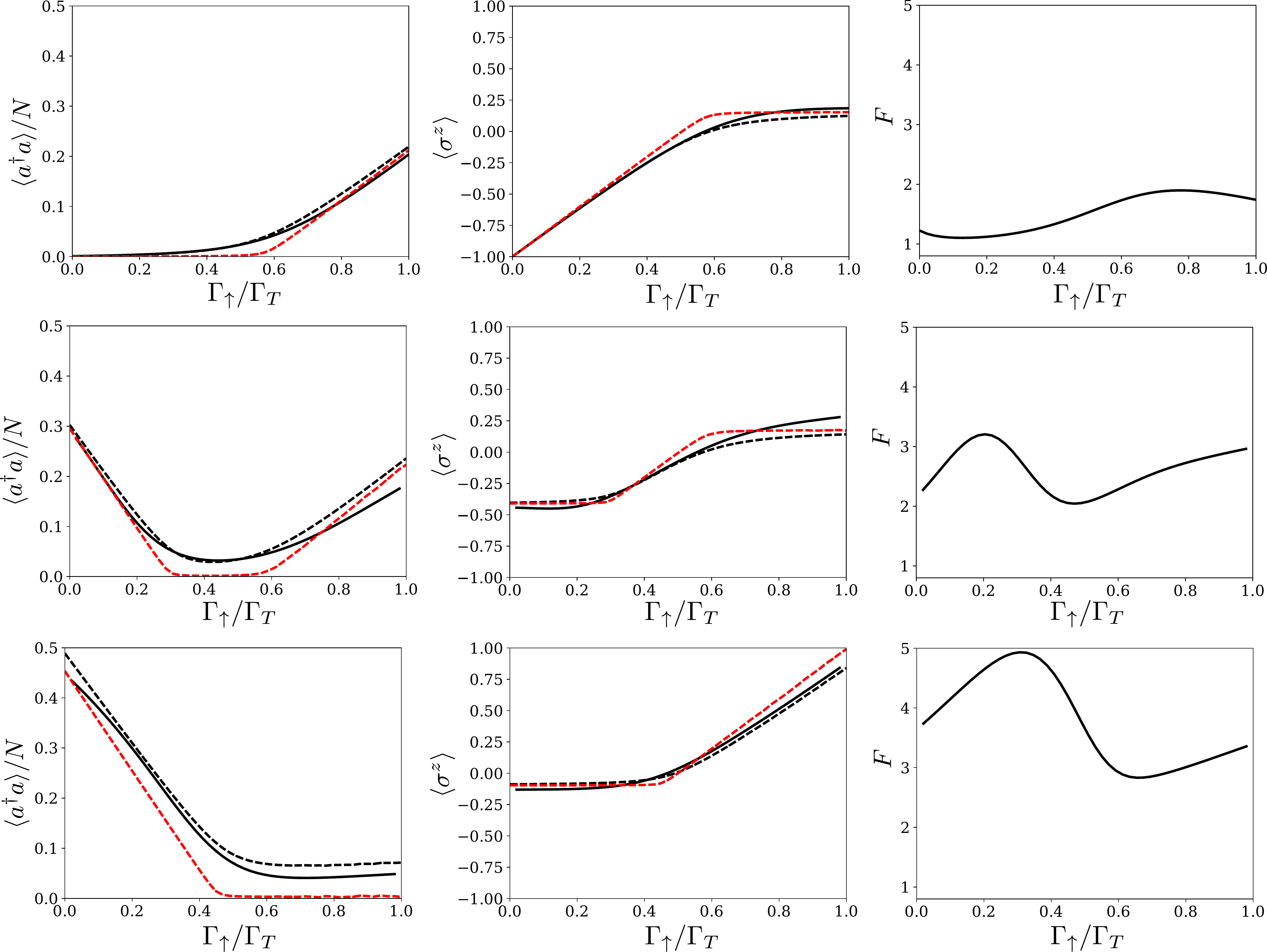}
 \caption{Photon number (left column), spin inversion (middle column) and Fano factor (right column) for three different values of $g'$. Top row: $g'=0.1g$, below the superradiant dome. Middle row $g'=g$ where both superradiance and lasing are observed for different values of $\gup$. Bottom row $g'=3g$ above the lasing region. The solid black lines are exact numerical results for $N=25$, the dashed black lines show the cumulant expansion for $N=25$ while the dashed red lines are the cumulant expansion close to the thermodynamic limit, $N=500$. All other parameters are the same as in Fig.~\ref{fig:MF}. \label{fig:compare_cum_gup}}
\end{figure}

To find exact numerical solutions of this system we can make use of the permutation symmetry of the individual density matrix elements. This allows us to find exact solutions at intermediate $N$. A full description of the method we use can be found in Ref.~\cite{Kirton2017} while the code can be found at~\cite{Kirton2017a}. Another library which implements the same algorithm has also recently been released~\cite{Gegg2017}.
Similar techniques have been employed to study spin ensembles~\cite{Chase08}, simple lasing models~\cite{Xu13}, coherent surface plasmons~\cite{Richter2015}, the competition between collective and individual decay channels~\cite{Damanet2016}, the behaviour of an ensemble of Rydberg polaritons~\cite{Gong2016}, equilibrium properties of a model with a larger local Hilbert space~\cite{Zeb2016},  subradiant states in the Dicke model~\cite{Gegg2017a} and to explore the effect of individual losses on transient superradiant emission~\cite{Shammah2017}.   
In this section, we use these exact solutions both to check the validity of the cumulant method presented above and calculate higher order moments of e.g.\ the photon distribution which are neglected in the cumulant calculation.  Indeed, from the exact results we can calculate not only moments, but the full probability distribution of the photon number.

Results comparing exact numerics to the cumulant expansion are shown in Fig.~\ref{fig:compare_cum_gup}. The top row shows a sweep through the lasing region (below the SR dome), the middle column goes through both the SR and lasing regions while the top row shows a sweep above the lasing region. 
The first two columns  show the reduced photon number $\ada/N$ the spin inversion level $\average{\sigma^z}$, which allow comparison to the mean-field and cumulant approximations.
We see that the match between the cumulant equations and the exact solution is good across all parameters, even at the relatively small value of $N=25$ shown here which is far from the thermodynamic limit,  with only some small deviations at large pump strengths.

The third column of Fig.~\ref{fig:compare_cum_gup} shows the Fano factor of the photon distribution,
\begin{equation}
	F = \frac{\average{\adag  a \adag a} - \ada^2}{\ada}.
\end{equation}
Calculating this quantity requires evaluating moments beyond second order, so is not accessible from our mean-field or cumulant approaches.  The Fano factor is one possible measure of the dispersion (spread) of the photon number distribution. For any coherent (Possonian) state the Fano factor is 1. This means that in the \new{thermodynamic} limit of an ideal laser the Fano factor of the lasing state is unity both above and below threshold. Directly at threshold the fluctuations give an infinite value for $F$~\cite{Rice1994}. However, it is known that finite size effects and finite non-radiative loss (i.e.\ finite $\beta$ factor) smear out the peak in the Fano factor so that it can become a very broad feature~\cite{Rice1994}. This is the effect we see here, where in the lasing phase the Fano factor is elevated far above unity by a very broad feature. The peak associated with the transition to the superradiant state is however much narrower. This can be clearly seen in the Fano factor at $g'=g$ where the structure is made up from these two peaks. This also explains why the cumulant expansion is less accurate in the lasing phase than it is in the SR phase: the cumulant expansion only describes the behaviour of approximately \new{Gaussian} states which the lasing phase we see does not satisfy.

\begin{figure}
\centering
 \includegraphics[width=1.0\columnwidth]{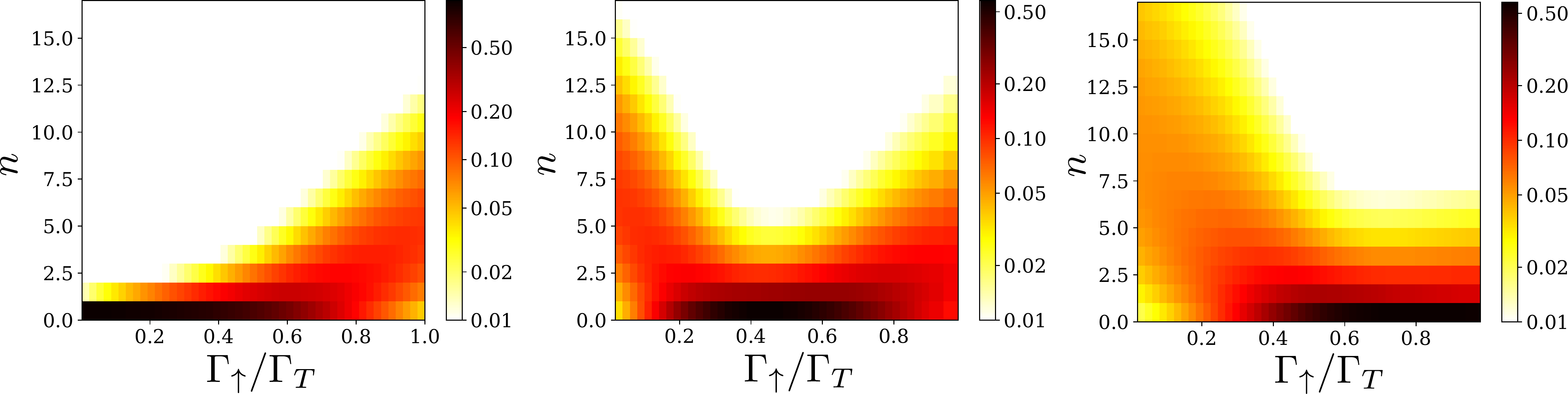}
 \caption{$P(n)$ distribution of the steady state photon density matrix for the same parameters as in Fig.~\ref{fig:compare_cum_gup} at $N=25$. The coherent driving in each panel is (a) $g'=0.1g$, (b) $g'=g$, (c) $g'=3g$.  \label{fig:Pn_gup}}
\end{figure}

In Fig.~\ref{fig:Pn_gup} we show the full probability distribution of the photon mode, $P(n)$, for the same sets of parameters as in Fig.~\ref{fig:compare_cum_gup}, i.e. with $N=25$. This allows us to see more clearly exactly what happens to the photon distribution as we go through the various thresholds. In the lasing regime as the pump power is increased the distribution moves to larger and larger $n$. The Fano factor does not drop down to 1 in this regime since for the parameters chosen here the peak of $P(n)$ only reaches $n=4$ at the largest pump strength. The opposite effect is seen in the SR regime where increasing pumping kills the coherent state. In the crossover regime we see both effects: At small pumping the SR phase is present which is killed by the incoherent drive, but at larger pumping a coherent state again appears as the system undergoes a transition to a lasing state.

\section{Two-time correlations and emission spectrum}
\label{sec:spectrum}

\begin{figure}
\centering{
 \includegraphics[width=0.8\columnwidth]{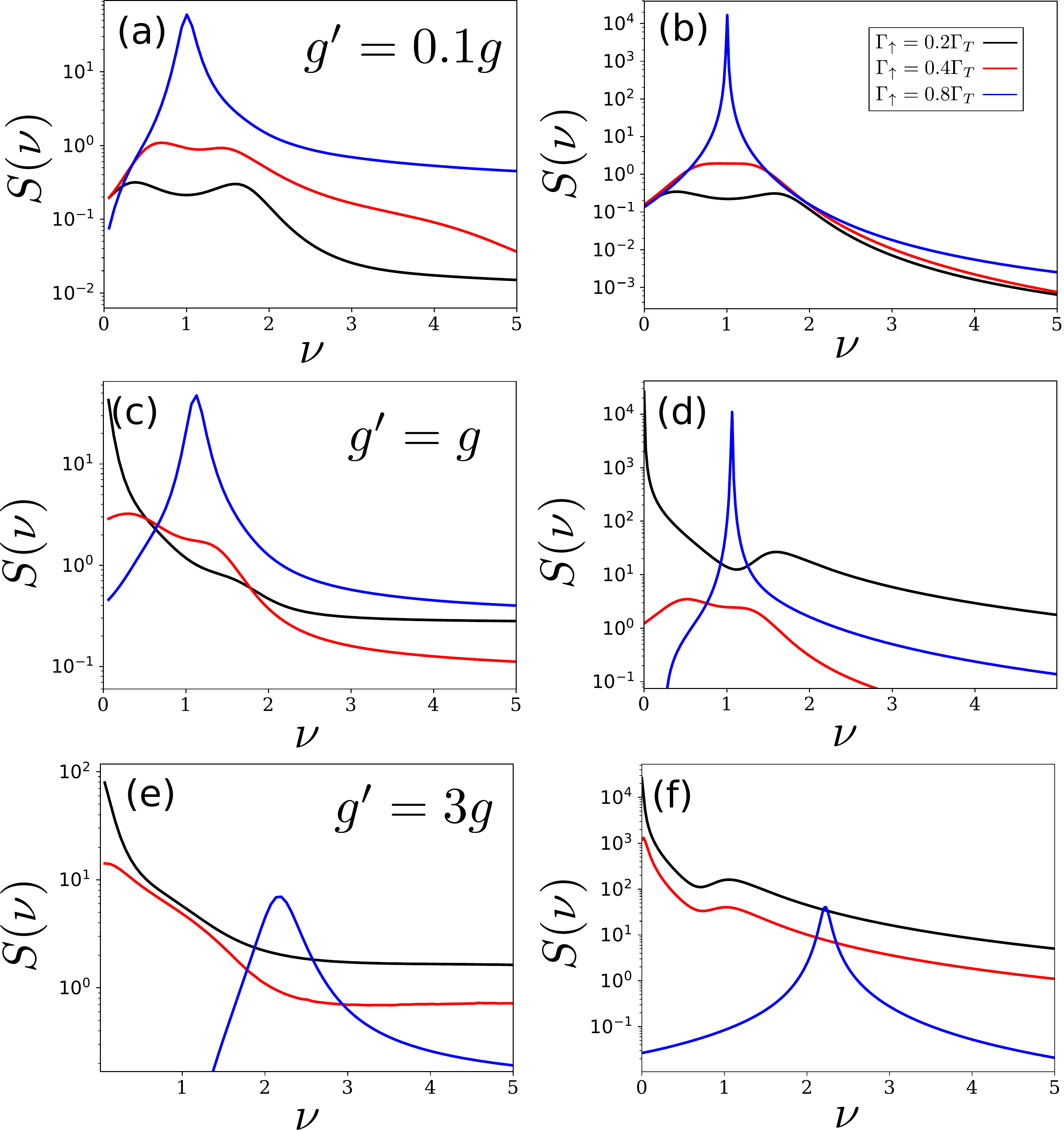}}
 \caption{Spectrum, $S(\nu)$, at various points in the phase diagram. The left column are exact results for $N=25$ while the right-hand column show the results from the cumulant expansion at $N=500$. The top row (a)-(b) have $g'=0.1g$, the middle row (c)-(d) is at $g'=g$, while the bottom row (e)-(f) have $g'=3g$. The pumping strengths in each case are, for the black curves $\gup=0.2\gdn$, red $\gup=0.4\gdn$, and blue $\gup=0.8\gdn$. Other parameters are the same as in Fig.~\ref{fig:MF}. \label{fig:spec}}
\end{figure}

The emission spectra of light-matter systems can reveal subtle features of the dynamics not available from the steady state behaviour.  A well known example of this is the Mollow triplet, seen in the fluorescence from a resonantly driven two-level system~\cite{Scully1997}. By examining the spectrum of the light which escapes the cavity we may be able to see the difference between the SR and lasing phases. 

The simplest quantity to look at, accessible from both our exact numerics and the cumulant expansion is the emission spectrum which is the Fourier transform of the $g^{(1)}(t)$ correlation function,
\begin{equation}
	S(\nu) = \int_{-\infty}^\infty \average{\adag(t)a(0)}\text{e}^{i\nu t} \,dt.
\end{equation}
We can obtain this quantity from our exact numerics using the quantum regression theorem. We find the steady state $\rho_{ss}$ as before and then initialise using the state $a\rho_{ss}$. The correlation function is then given by the time evolution of $\adag$ from this state~\cite{Scully1997}. The spectrum is then obtained by taking a Fourier transform. 

In the large $N$ limit we may also use the second order cumulant equations and quantum regression theorem to calculate the same quantity. To do this we must solve the matrix equation
\begin{equation}
	\frac{d}{dt}\mathbf{C}(t) = \mathcal{M}\mathbf{C}(t) ,
\end{equation}
where the coupled set of correlation functions required are
\begin{equation}
	\mathbf{C}(t) =\begin{pmatrix}
	            	\average{\adag(t)a(0)} \\
	            	\average{a(t)a(0)} \\
	            	\average{\sigma^+(t)a(0)} \\
	            	\average{\sigma^-(t)a(0)}
	            \end{pmatrix},
\end{equation}
and the evolution matrix is given by
\begin{equation}
\mathcal{M} = 
\begin{pmatrix} 
	i\omega -\kappa/2 & 0 & ig & ig' \\ 0 & -i\omega -\kappa/2 & -ig' & -ig \\ -ig\average{\sigma^z}_{ss} & -ig'\average{\sigma^z}_{ss} & i\omega_0-\Gamma_T/2 & 0 \\ ig'\average{\sigma^z}_{ss} & ig\average{\sigma^z}_{ss} & 0 & 	-i\omega_0-\Gamma_T/2                                                                             \end{pmatrix}.
\end{equation}
A similar approach was taken in Ref.~\cite{Gong2016} to calculate correlation functions of an ensemble of Rydberg polaritons. Here $\sz_{ss}$ is the steady state value of the spin inversion and the initial condition for this problem is given by the relevant quantities from the second moment equations. 

These two methods then allow us to calculate the spectrum as shown in Fig.~\ref{fig:spec}. The simplest interpretation can be found by looking at the results of the cumulant expansion at $N=500$ (the right-hand column in Fig.~\ref{fig:spec}). When the counter-rotating term is weak (i.e. $g'=0.1g$), and the incoherent pump is strong, the spectrum  has a single peak at the cavity frequency which gets more intense as the pumping is increased. This monochromatic emission is typical of a simple laser.  We may also note that the emergence of this non-zero single frequency in the lasing state is directly related to the Hopf bifurcation that leads to the lasing state.
 In the crossover region, $g'=g$ at low pump powers (in the SR dome) we see a  triplet like structure, reminiscent of the Mollow triplet,  with the sideband at a frequency which is not simply set by the cavity. As the incoherent pumping is increased the spectrum changes to look more that seen for the case of the laser. At large $g'=3g$ (above the lasing region in Fig.~\ref{fig:MF}) we see a very similar structure but the residual peak at the cavity frequency is much weaker. This peak is still present since the cumulant calculations are done at finite $N$ and the parameters chosen are relatively close to the lasing region of the phase diagram.  The $N=25$ results form the exact numerics in the left hand column of Fig.~\ref{fig:spec} show a similar qualitative behaviour but the relatively small value of $N$ means that the features are less easy to distinguish.

\begin{figure}
\centering{
 \includegraphics[width=0.95\columnwidth]{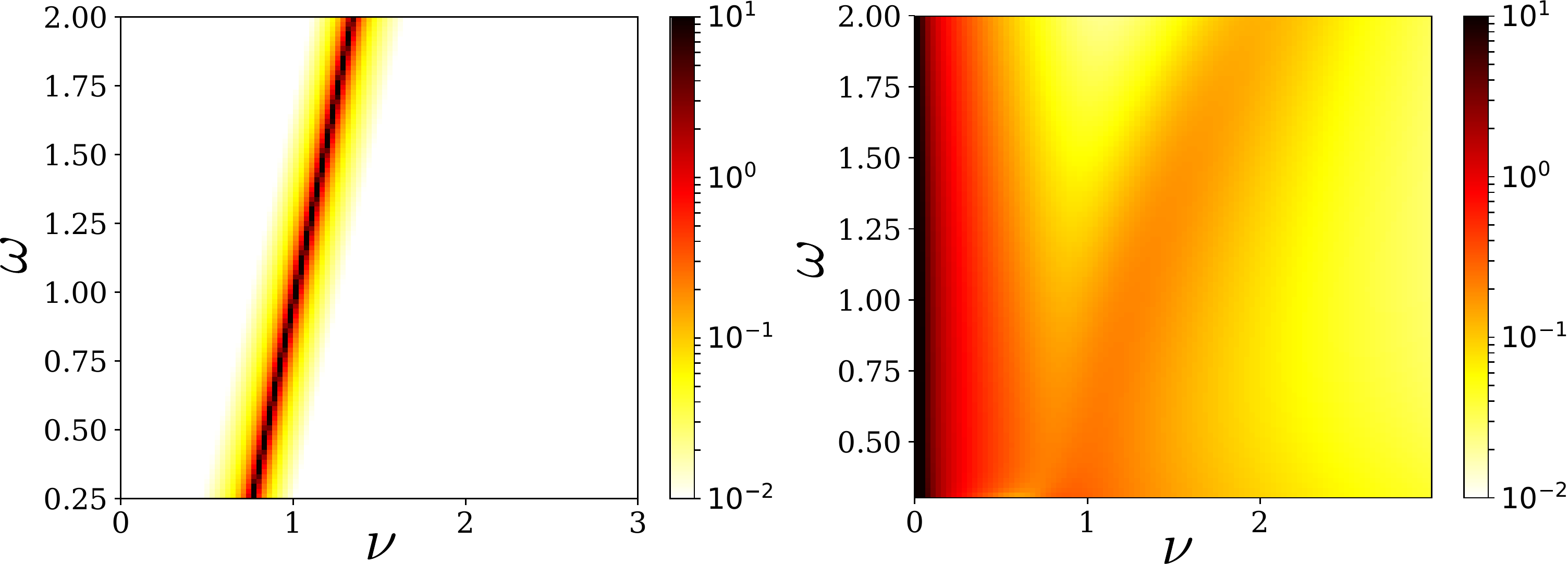}}
 \caption{Spectrum, $S(\nu)$, as a function of photon frequency $\omega$ (a) in the lasing regime $g'=0.1g$, $\gup=0.8\Gamma_T$, (b)  in the SR phase $g'=2g$, $\gup=0.2\Gamma_T$. All other parameters as in Fig.~\ref{fig:MF}. \label{fig:spec_om}}
\end{figure}

In contrast to the steady state measurements discussed in previous sections,
we see here that the emission spectrum of the cavity provides an easy way to identify the differences between the SR and lasing regimes.  This difference can be particularly clearly seen in the spectrum as a function of cavity frequency $\omega$. In Fig.~\ref{fig:spec_om} we show this calculated both in the lasing phase and the SR phase. In the laser the location of the peak is approximately linear in the cavity frequency consistent with the interpretation of this being a weak light-matter coupling effect, while in the SR phase the sideband has a more complex dependence on the detuning.

\section{Blue detuning and inverted states}
\label{sec:inverted}

\begin{figure}
\centering{ \includegraphics[width=1.0\columnwidth]{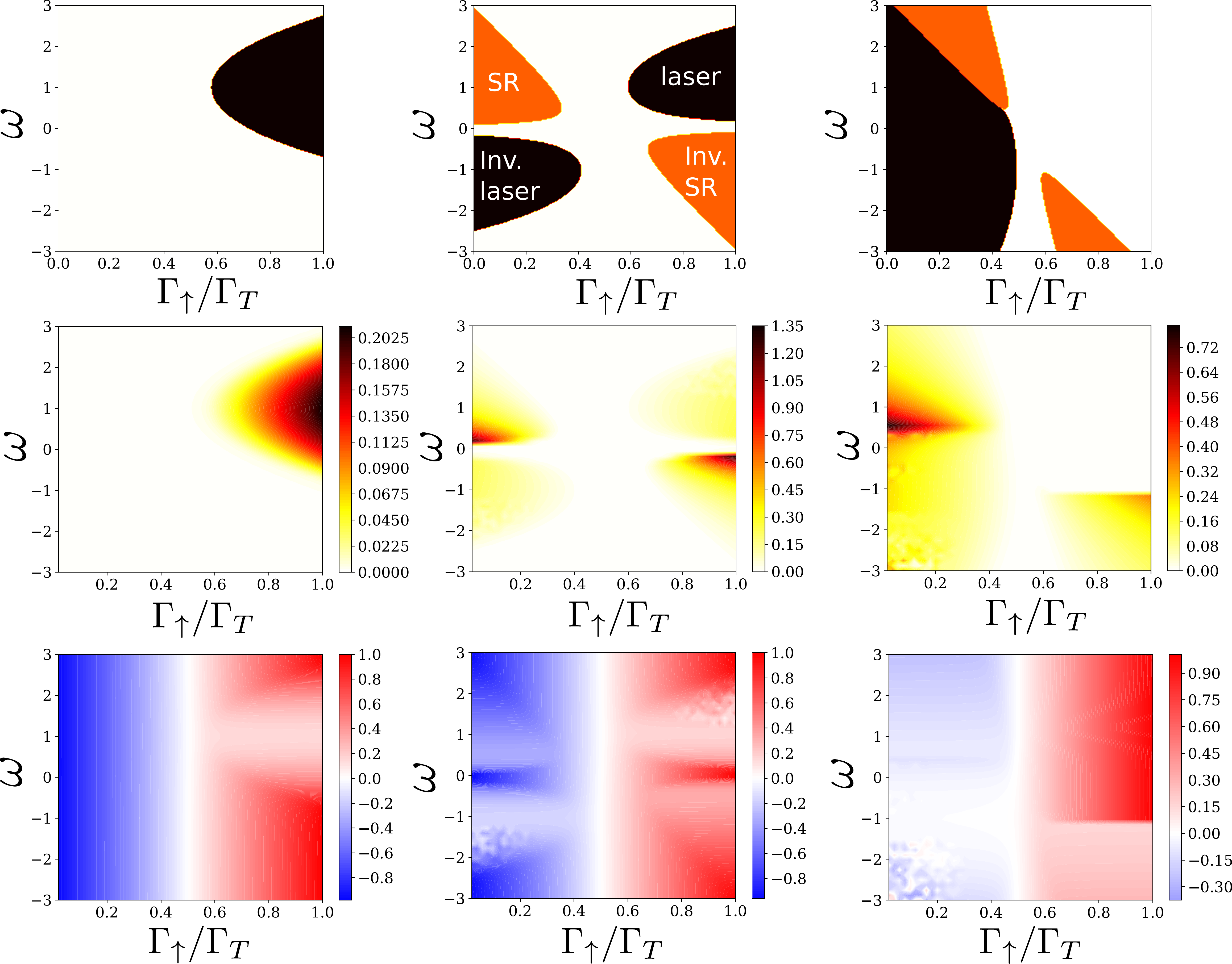}
 \caption{\label{fig:vs_om} Phase diagram vs $\omega$ at different values of $g'$. Left column $g'=0.1g$, centre $g=g'$ and right $g'=3g$. The top row shows the mean field phase diagram, as in Fig.~\ref{fig:MF}, where again orange indicates a single unstable mode, while black indicates a pair of unstable modes. The middle and bottom rows show the photon number and spin inversion, $\sz$, from the cumulant expansion, as in Fig.~\ref{fig:Gup_gp}. All other parameters are the same as in Fig.~\ref{fig:MF}}}
\end{figure}

In the model we consider, the effective cavity frequency, $\omega$, is actually the detuning between the cavity and external pump frequency (see the level diagram in Fig.~\ref{fig:schematic}).  As such, it is reasonable to ask how the behaviour changes when this detuning is negative, i.e. when the pump is blue detuned from the cavity mode.  From previous work, we know that if only photon loss is present then the normal state in this parameter range is inverted~\cite{Bhaseen2012}. In this section we therefore discuss what happens to this inverted regime when there is incoherent pumping and decay of the spins.

In Fig.~\ref{fig:vs_om} we show the phase diagram as a function of both the incoherent pumping rate and photon detuning for various values of $g'$. If $g'\ll g$ then the only phase that is present is the normal lasing state which appears at large $\gup$ and for resonant frequencies $\omega\approx \omega_0$. In the middle column of Fig.~\ref{fig:vs_om} we see that when $g=g'$ more phases appear, both above and below inversion. When $\omega>0$ we see the normal SR and lasing phases at small and large $\gup$ respectively. At negative $\omega$ we see two very similar phases but at opposite sides of the phase diagram. To understand these we note that the model has a duality under the replacements $\omega\leftrightarrow -\omega$, $g\leftrightarrow g'$, $\gup \leftrightarrow \gdn$: such a transformation leaves the steady states unchanged up to an inversion of $\sigma^z$.  This then means that the two phases seen at negative $\omega$ are the same as those when $\omega$ is positive but with photon creation and annihilation swapped, i.e.\ the inverted lasing phase at small $\gup$  is where the dominant process is photon and spin excitation creation via the term $g'\adag \sigma^+$ in the Hamiltonian. This state is similar to one discussed in Refs.~\cite{Kapit2014, Hafezi2015} where a parametric drive is used to engineer similar behaviour.

In this region of parameter space we  see some artefacts in the results found by time evolving the cumulant expansion (most visible in the spin dynamics). These artefacts reflect the fact the photon number and spin inversion do not reach a steady state, but instead one has limit cycles and chaotic dynamics. We discuss this behaviour more below. As $g'$ is increased further we see that the phases below inversion --- the normal SR and inverted lasing phases --- coalesce, while the two phases at large $\gup$ are suppressed. Indeed, by the point $g'=3g$, shown in the right hand column of Fig.~\ref{fig:vs_om}, the standard lasing phase has completely vanished.  Although the SR and lasing phases join continuously to each other at $g'=3g$, one can nonetheless distinguish two distinct behaviours on crossing the phase boundary: it remains the case that one can distinguish whether a single eigenvalue or a pair of eigenvalues becomes unstable.  At this largest value of $g'$, the absence of a steady state photon number extends over a wider portion of the phase diagram at negative $\omega$. 

Careful comparison  of the mean field and cumulant expansion phase diagrams reveal that close to the upper boundary of the inverted SR phase there is a region where mean-field stability analysis shows that the normal state is stable and yet the cumulant equations show a macroscopic photon population. The equations in this regime are bistable: while the normal state is stable, there is also a stable SR solution to the equations. Hence we find that in this region of the phase diagram the results of the cumulant equations depend on the initial conditions.

\subsection{Evolution of chaotic attractors}

\begin{figure}
 \centering{\includegraphics[width=0.9\columnwidth]{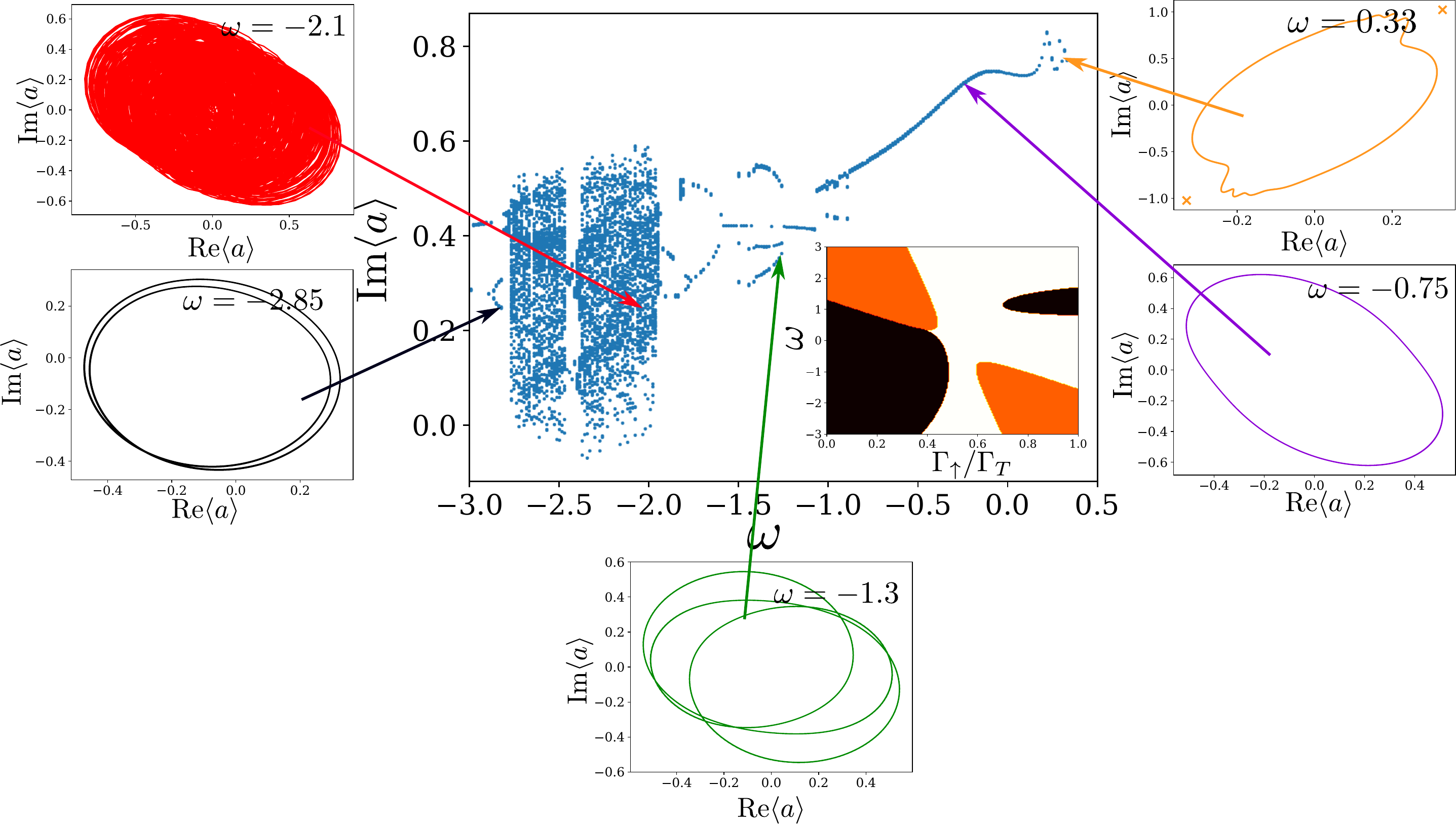}
 \caption{\label{fig:bifurcation}  Bifurcation diagram vs cavity frequency $\omega$ calculated from the mean field equations. The main plot shows the values of $\Im\average{a}$ recorded when $\Re\average{a}$ crosses the axis,
for $g'=2.3g, \gup=0.01$, while the inset shows the mean field stability phase diagram at this value.  The subplots show the long time dynamics of the photon at the frequencies indicated. All other parameters as in Fig.~\ref{fig:MF}.}}
\end{figure}

To characterise the non-steady-state behaviour in more detail we look at the dynamics of the mean field equations as the frequency of the photon mode is changed. These results are shown in Fig.~\ref{fig:bifurcation}. We plot the dynamics of the photon mode using an initial condition with finite $\average{a}$ to break the symmetry (the normal state with $\average{a}=0$ is always a solution to the MF equations, but in the SR phase this solution is unstable), and show the behaviour after waiting sufficiently long that any transient behaviour has vanished. To plot the bifurcation diagram~\cite{sprott2003} in the centre of Fig.~\ref{fig:bifurcation} we plot the value of $\Im\average{a}$ each time the photon crosses the $\Re\average{a}=0$ line from right to left. 

In the region where $\omega$ is closest to zero we see that the dynamics are regular, the photon amplitude undergoes a simple periodic limit cycle which results in a single point in the bifurcation diagram. As the frequency is decreased (away from 0) the system goes through a sequence of transitions to more complex (but still regular) orbits which eventually lead to the chaotic region at around $\omega=-2.0$. At larger negative detunings again we find that the dynamics become regular. At positive detunings, e.g. $\omega=0.33$, the limit cycle transitions towards a fixed point: as we approach this, the limit cycle distorts, and the time evolution pauses near the points that ultimately become the fixed point values.  These fixed points are indicated by crosses in this panel. The occurrence of chaotic behaviour within the lasing region is not too surprising since it is known that in certain limits the Maxwell-Bloch equations can be mapped to the Lorentz equations~\cite{Haken1975a}.

\section{Conclusions}
\label{sec:conclusions}

We have examined in detail the steady state phase diagram of a model which is able to continuously cross over from standard two-level lasing to Dicke superradiance by changing the balance between coherent and incoherent pumping processes.  We have seen that when the photon energy is positive these two phases are distinct, the lasing phase only exists when the spins are inverted and the SR phase when the spins are not inverted. 

By using a combination of approaches: mean-field theory valid in the thermodynamic limit, a cumulant expansion which gives access to the large $N$ asymptotics and exact numerics which reach intermediate $N$ we have shown that while the steady state photon number is similar in the two phases the spectral properties of the emission from the cavity are very different.

The phase diagram as a function of the effective photon energy $\omega$ reveals two more phases when $\omega<0$. An inverted lasing phase at weak incoherent pump strengths and an inverted SR phase at large incoherent pump. This also revealed the evolution of the dynamics through a period doubling route to a chaotic attractor.

Future studies could examine the role of the non-linearity which arises in the Hamiltonian due to the ac Stark shift of the cavity mode~\cite{Keeling2010, Bhaseen2012, Gelhausen2017}. This is known to generate limit cycles in the mean field dynamics, it would be interesting to see what effect the incoherent processes considered here have on these phases.

\ack 
P.K.\ acknowledges support from EPSRC (EP/M010910/1).  J.K.\ acknowledges support from EPSRC programs “TOPNES” (EP/I031014/1) and ``Hybrid Polaritonics'' (EP/M025330/1).

\section*{References}


\providecommand{\newblock}{}

\end{document}